# External-Field-Free Spin Hall Switching of Perpendicular Magnetic Nanopillar with a Dipole-Coupled Composite Structure

Zhengyang Zhao[†], Angeline K. Smith[†], Mahdi Jamali, and Jian-Ping Wang [*]

Department of Electrical and Computer Engineering, University of Minnesota, 4-174 200 Union Street SE, Minneapolis, MN 55455, USA

Spin Hall effect (SHE) induced reversal of perpendicular magnetization has attracted significant interest, due to its potential to lead to low power memory and logic devices. However, the switching requires an assisted in-plane magnetic field, which hampers its practical applications. Here, we introduce a novel approach for external-field-free spin Hall switching of a perpendicular nanomagnet by utilizing a local dipolar field arising from an adjacent in-plane magnetic layer. Robust switching of perpendicular CoFeB nanopillars in a dipole-coupled composite stack is experimentally demonstrated in the absence of any external magnetic field, in consistent with the results of micromagnetic simulation. Large in-plane compensation field of about 135 Oe and out-of-plane loop shift of about 45 Oe / $10^7$ A·cm$^{-2}$ are obtained in the nanopillar devices with composite structure. By performing micromagnetic simulations, we confirm the composite external-field-free switching strategy can also work for a $10 \times 10$ nm$^2$ circular pillar. Compared with other proposed methods for external-field-free spin Hall switching of perpendicular magnetization, the dipole-coupled composite structure is compatible with a wide range of spin Hall systems and perpendicular magnetic tunnel junctions, paving the way towards practical SHE-based MRAM and logic applications.

*Corresponding author. Tel: (612) 625-9509. E-mail:   jpwang@umn.edu
† These authors contributed equally to this work.

# Introduction

Manipulation of the magnetization through spin-orbit torques (SOT) arising from spin Hall effect (SHE)[1–5] has attracted immense attention due to its potential applications in magnetic random-access-memory (MRAM) and magnetic logic devices. Compared with conventional spin-transfer torques (STT)[6,7] generated by a current passing across an ultrathin tunnel barrier in a magnetic tunnel junction (MTJ), SHE-induced writing mechanism allows the implementation of a 3-terminal memory cell[2,8–10], where the write current is injected in the spin Hall (SH) channel instead of the tunnel barrier thus it increases the reliability and endurance of the memory cell. Moreover, utilizing giant spin Hall efficiency materials[4,11–14] promises a path for drastic reduction of the writing current.

Perpendicular MTJ (p-MTJ) composed of storage layer(s) with perpendicular magnetic anisotropy (PMA) is preferred over the in-plane MTJ (i-MTJ) in MRAM devices, because p-MTJ offers a higher storage density while maintaining the thermal stability and operates at lower power[7,15]. Since SHE-induced torques are symmetric in respect to the perpendicular magnetization orientation, an additional torque is require to break this symmetry. Traditionally, this has been accomplished by applying an external magnetic field parallel to the current channel[1–4,16–19], which imposes a limitation for the applications. Recently, attempts have been made to achieve field-free SH switching with two alternative symmetry-breaking mechanisms. The first approach is based on asymmetric geometric structures[20–25], which would be difficult to manufacture for high areal density memory arrays. The second proposal is based on having an anti-ferromagnetic (AFM) material in contact with the perpendicular magnet to provide an in-plane exchange bias field $H_{ex}$ thus breaking the switching symmetry[26–32]. This approach provides a path towards SHE based memory devices if the AFM layer also serves as the spin Hall channel[28–30,32]. However, it restricts the utilization of a wide range of more efficient SH materials such as topological insulators[4,11–14]. Additionally, the challenges of obtaining sufficient in-plane exchange bias field[27–32], the existence of multi-domain states in the AFM layer[27,29,31,32], as well as the low blocking temperature in the AFM-SH systems[31] make its implementation for memories not so straightforward. Finally yet importantly, all the experimental results based on having AFM layer are performed with Hall bar devices (see Table 1). Robust external-field-free switching of nano-scaled magnetic pillar is yet to be demonstrated.

Here we propose a new strategy to realize external-field-free SH switching of a perpendicular magnet by using a dipole-coupled composite structure, where the lateral symmetry of the perpendicular magnet is broken by the dipolar field, or stray field, generated from a top in-plane magnetic layer (as illustrated in Fig. 1a). Robust full switching of perpendicular Ta/CoFeB/MgO nanopillar in the composite stack is demonstrated in the absence of any external field. The preferred orientation of the PMA nanomagnet is determined by both the direction of channel current and the direction of the in-plane magnetization. Additionally, the switching process of a $300 \times 150$ nm$^2$ nanopillar is studied using micromagnetic simulations involving the non-uniform stray field. The field-free switching of a $10 \times 10$ nm$^2$ composite pillar is also demonstrated by simulation, confirming the scalability of the composite structure. Compared with previously proposed methods, our approach based on the dipole-coupling is well compatible with various giant spin Hall angle materials and readily applicable for SOT-based memory and logic devices.

## Results

**Concept of the p-MTJ integratable composite structure.** Conventional PMA spin Hall device consists of a perpendicular magnetic layer (PL) developed on a spin Hall channel layer. In our proposed composite structure, an additional in-plane magnetic layer (IPL) is incorporated on top of the PL structure which may be separated from the PMA layer by a nonmagnetic metal/oxide layer. Both magnetic layers are patterned into elliptical nanopillars on top of the Ta SH channels, as illustrated in Fig. 1a. Since the long axis of the pillar is along the SH channel in the $y$-direction, the IPL generates a dipolar magnetic field that is mainly along the *y*-axis, parallel or anti-parallel with the direction of the channel current depending on the IPL magnetic orientation. The presence of the *y*-component of the stray field breaks the lateral symmetry of the PL, leading to deterministic switching of the perpendicular magnetization. The shape of pillar could also be circular if the orientation of the IPL magnet is set utilizing an AFM layer on top. This composite structure could be incorporated into the existent p-MTJ with a minimal modification of the stack for SOT based applications, as demonstrated in Fig. 1d.

**Preparation of nanopillar SH devices.** The stack of the composite structure consists of Ta (5)/CoFeB (1.2)/MgO (2)/CoFeB (3)/MgO (2)/Ta (5) (from bottom to top, with thicknesses in nanometers), where the CoFeB (1.2 nm) layer has a perpendicular anisotropy and the CoFeB (3 nm) layer has an in-plane anisotropy. An additional sample with the stack structure of Ta (5)/CoFeB (1.2)/MgO (2)/Ta (5) is prepared as a reference. Both structures are patterned into elliptical nanopillars on top of Ta Hall bars, as shown in the scanning electron micrograph in Fig. 1b. To characterize the magnetization direction of the perpendicular CoFeB layer, the Hall resistance $R_H$ due to the anomalous Hall effect (AHE) is measured at a given channel current $I_{ch}$. All the experimental results presented in the main text are obtained from devices with channel width of 500 nm and nanopillars dimension of $285 \times 95$ nm$^2$ (aspect ratio 3:1). The results for nanopillars of $300 \times 120$ nm$^2$ (2.5:1) and $300 \times 150$ nm$^2$ (2:1) are included in the Supplementary Materials.

The perpendicular anisotropy of the perpendicular CoFeB layers is verified by carrying the vibrating sample magnetometer (VSM) measurement (as seen in Fig. 1c). The effective perpendicular anisotropy field is determined to be $H_k \approx 0.35$ T for both the composite and the reference samples (see Methods), indicating the PMA of the PL in the composite structure is barely affected by the IPL layer on top.

**$R_H$-$H_z$ loop shift of the composite nanopillar.** Hall resistance $R_H$ reflecting the magnetization orientation of the PL nanomagnet is first examined by scanning the out-of-plane field $H_z$, at different channel currents $I_{ch}$. Figure 2a shows the $R_H$-$H_z$ loops for the composite nanopillar, obtained at $I_{ch} = \pm 0.3$ mA, $\pm 0.4$ mA, and $\pm 0.5$ mA. Note positive $I_{ch}$ is along $+y$ direction. The IPL magnetization is set to $-y$ direction by an external in-plane field before the measurement. With increasing $|I_{ch}|$, the $R_H$-$H_z$ loops shrink and the coercive field of the loop ($H_c$) gradually decreases due to the increase of SOT as well as Joule heating, which is the common feature of PMA spin Hall devices[19,33]. In addition, the center of the loop is shifted toward left ($I_{ch} > 0$) or right ($I_{ch} < 0$) at large $|I_{ch}|$. At $I_{ch} = \pm 0.5$ mA, the separation between the two loops is $H_{shift}$ ($-0.5$ mA) $- H_{shift}$ ($+0.5$ mA) $\approx 187$ Oe. The full dependence of $H_c$ and $H_{shift}$ on channel current $I_{ch}$ is presented in Fig. 2c. On the other hand, if the IPL magnetization is pointing to $+y$ direction, the $R_H$-$H_z$ loop would shift to the right for $I_{ch} > 0$ and to the left for $I_{ch} < 0$, as seen in Fig. 2d. In comparison, the reference sample shows no obvious shift of loop (as shown in Fig. 2b). The four-fold symmetry in the $H_c$-$I_{ch}$ phase diagram of the Ta/CoFeB/MgO reference device (Fig. 2b) becomes two-fold symmetry with the IPL added on top of the stack (Fig. 2c & d), indicating $I_{ch}$ with opposite signs favors opposite orientations of PL in the composite sample.

Such breaking of the four-fold symmetry in the $H_c$-$I_{ch}$ diagram has also been observed in some of the recent external-field-free SH switching devices, induced either by an PMA gradient[20–22,25], or by an in-plane exchange bias field[26,29,32]. Here in the dipole-coupled composited structure, the breaking of symmetry is induced by the stray field from the IPL. When IPL is set to $-y$ direction (Fig. 2a & c), the direction of its stray field experienced by the PL is mainly along $+y$. As a result, up magnetization of PL is preferable at $I_{ch} > 0$ and down magnetization of PL is preferable at $I_{ch} < 0$. If IPL is set to $+y$ direction (Fig. 2d), the preferred state becomes opposite. The strength of the inversion asymmetry can be characterized with the strength of the out-of-plane effective field per current density, or $|H_{shift}/J_{ch}|$[20,25]. For our devices, $|H_{shift}/J_{ch}|$ is about 45 Oe / $10^7$ A·cm$^{-2}$ (obtained from Fig. 2c & d and Fig. S2 in the Supplementary Materials), comparable to the PMA gradient devices[20] and greater than the exchange-coupled devices[26,29] as summarized in Table 1. Additionally, for $|I_{ch}| > 0.35$ mA (the regions marked with green/yellow in Fig. 2c & d), the $R_H$-$H_z$ loop is completely shifted to the negative or positive field region, suggesting that only one orientation of PL is stable at $H = 0$. This implies that the external-field-free switching could be obtained with $|I_{ch}| > 0.35$ mA.

**Switching of PL nanomagnet measured with various external in-plane field $H_y$.** The influence of the IPL biasing layer is then studied by sweeping the external field $H_y$ in the presence of various channel currents, as shown in Fig. 3a & b. Under a possitive current $I_{ch} = +0.35$ mA, PL has a up magnetization when $H_y > 0$ (as suggested by the positive $R_H$ values), and a down magnetization when $H_y < 0$ (as suggested by the negative $R_H$ values). This can be explained by the symmetry breaking due to the application of the external field $H_y$[3]. However, the switching of PL occurs before $H_y$ is swept across the zero point (as shown in Fig. 3b), which is different from traditional unbiased SH devices[18]. Additionally, under a larger current of $I_{ch} = +0.55$ mA, it's interesting to note that multiple switching events occur within the external field range of −150 Oe ~ +150 Oe (as shown in Fig. 3a), which hasn't been observed in previous perpendicular SH devices.

The unique behaviors of the composite nanopillar can be well explained as a result of the competition between the stray field $H_{stray}$ of IPL and the external field $H_y$. For better understanding of the switching behaviors, $R_H$-$I_{ch}$ loops are measured with various $H_y$ gradually changing from −270 Oe to +270 Oe, as presented in Fig. 3c. Five critical points A-E labelled in Fig. 3a are also pointed out in Fig. 3c, with the magnetization configuration of each state plotted on the right side of the $R_H$-$I_{ch}$ loops. For the sack of simplicity, it is assumed the stray field of IPL is a uniform field along $y(-y)$ direction. In real case, $H_{stray}$ is not uniform accompanying with an out-of-plane component as well that is further discussed later in the text.

Initially upon the application of $H_y = -270$ Oe, the IPL is aligned along $H_y$ pointing to $-y$ direction (point A). As a result, the stray field at PL location is mainly along $+y$ direction thus opposing the external field. The total field experienced by PL is $H_{tot} = H_y + H_{stray}$. Considering $|H_y| > |H_{stray}|$, $H_{tot}$ is dominated by $H_y$ pointing to $-y$ direction. Therefore, the $R_H$-$I_{ch}$ loop is clockwise (CW)[19]. Upon decreasing $H_y$ down to $-45$ Oe (point B), the $R_H$-$I_{ch}$ loop becomes counter-clockwise (CCW), suggesting $H_{stray}$ is dominant over $H_y$ thus $H_{tot}$ is along $+y$ direction. This corresponds to the switching between A and B in Fig. 3a. With further reduction of $H_y$ to 0 (point C), $H_{tot} = H_{stray}$ still points to $+y$, and the CCW loop keeps unchanged. When $H_y$ becomes $+45$ Oe, the polarity of $R_H$-$I_{ch}$ loop changes back to clockwise, corresponding to the sudden jump between C and D in Fig. 3a. This is due to the reversal of IPL and $H_{stray}$, since $H_y = +45$ Oe is larger than IPL's coercive field $H_c^{IPL}$ (note $H_c^{IPL}$ depends on both the aspect ratio of the nanopillar and the Joule heating induced by the channel current, as described in the Supplementary Materials). Finally, by increasing $H_y$ above $|H_{stray}|$, the third change of $R_H$-$I_{ch}$ loop polarity is observed, corresponding to the switch between D and E in Fig. 3a. Thus, the unique behaviors of the composite nanopillar under an in-plane external field result from the competition between $H_{stray}$ and $H_y$: under a large $|H_y|$ the favored orientation of PL is determined by the direction of $H_y$, similar with the case of traditional unbiased SH devices; under a small $|H_y|$ or without any external field, PL also has a preferred orientation which is determined by the direction of $H_{stray}$ instead of $H_y$.

We plot the variation of the critical current $I_c$ (the switching point of $R_H$-$I_{ch}$ loop) as a function of $H_y$, as shown in Fig. 3d-e. Compared with unbiased SH devices[19], the boundary between the clockwise and counter-clockwise loops shifts from zero to $-100$ Oe ($+100$ Oe) for the case of IPL // $-y$ (IPL // $+y$) owing to the existence of $H_{stray}$. For the nanopillars with aspect ratio of 2:1, this compensation point $H_{comp}$ even reaches up to 135 Oe (see Supplementary Materials), significantly overriding most of the exchange-coupled structures [27–30,32] as summarized in Table 1.

**Robust external-field-free switching of the composite nanopillar.** From Fig. 3c we have observed the switching of PL nanomagnet in the absence of any external field. Figure 4a & b show external-field-free switching loop with IPL // −y and IPL // +y, respectively. Consistent with previous analysis, $I_{ch}$ parallel with IPL (i.e. anti-parallel with $H_{stray}$) favors down magnetization of PL, while $I_{ch}$ anti-parallel with IPL (i.e. parallel with $H_{stray}$) favors up magnetization of PL. Figure 4c presents the robust external-field-free toggling of the nanopillar measured with a sequence of current pulses. The pulse amplitude is 0.6 mA and pulse width is 10 µs. Unlike the partial switching obtained in the exchange-coupled devices[27–29,31,32] which is attributed to the multi-domain state in the anti-ferromagnetic layer, the switching in our composite devices is abrupt and completed, suggesting the dipole-coupled devices would be more desirable for memory applications.

In addition to the implementation of memory devices as illustrated in Fig. 1d, the composite structure also allows one to implement logic gates by utilizing the additional degree-of-freedom, namely the IPL's magnetization orientation. Schematic in Fig. 4d shows an XOR gate based on this concept. Different from the memory device shown in Fig. 1d where the IPL is pinned, for the XOR gate the magnetization of IPL is controlled by the spin current in a top SH channel. Then the final state of the p-MTJ is determined by both the bottom channel current (Input Y) and top channel current (Input X). This provides a pathway towards the logic-in-memory applications[34,35].

## Discussion

**Analysis of switching process considering the non-uniform profile of stray field.** In above sections the experimental results are qualitatively explained by assuming $H_{stray}$ as a uniform field across the PL nanomagnet. In reality, however, the stray field has a non-uniform profile, as shown in Fig. 5a & b. Figure 5a illustrates the variation of *y*-component ($H_{stray}^{y}$) and *z*-component ($H_{stray}^{z}$) of the stray field at different locations of PL (with IPL pointing to −*y*), and Fig. 5b plots the profile of $H_{stray}^{y}$ and $H_{stray}^{z}$ based on micromagnetic calculations. It can be seen the direction of $H_{stray}^{z}$ reverses at the center of the pillar, and both $H_{stray}^{y}$ and $H_{stray}^{z}$ get stronger at each end of the elliptical pillar's long axis. Additionally, it has been reported for nanomagnets with the dimension of a few hundred nanometers, both the SH switching[36–39] and dipole-field induced switching[40] processes contains two steps: domain nucleation and domain wall propagation. Therefore, for a thorough understanding of the SH switching process, we have performed micromagnetic simulations trying to reproduce our experimental field-free switching results.

The first column in Fig. 5c shows the simulated evolution of PL magnetization under the non-uniform stray field (no external field is applied). Initially the PL magnet is pointing down representing by blue pixels. Upon the application of the channel current, a reversed domain is first nucleated near the +*y* end of the pillar followed by the expansion of the domain, since both $H_{stray}^{y}$ pointing to +*y* and $H_{stray}^{z}$ pointing to +*z* favors the reversal. However, if only considering the SHE, the reversed domain is hardly to expand to the whole pillar, because $H_{stray}^{z}$ pointing to −*z* at the other end of the pillar obstructs the magnetization reversal. This does not agree with the experiments where complete switching has been obtained. To solve this discrepancy, we have considered two additional mechanisms that may assist the expansion of the reversed domain: (a) domain wall propagation driven by conventional spin-transfer torque (STT) in the PL layer (as seen in the second column in Fig. 5c); and (b) domain wall propagation controlled by SHE in the presence of Dzyaloshinskii-Moriya interaction (DMI)[18,41–43]. Simulations shows both mechanisms can drive the expansion of the reversed domain to the entire pillar. For our case, the full switching of the nanomagnet is more likely via the simultaneous action of spin orbit torque and DMI, since STT-driven domain wall motion tends to take place in structures with thinner underlayer[24].

If the nanopillar is shrunk down to below 40 nm, the magnetization switching mode would change from nucleation type to single domain type[37]. Then the non-uniform profile of the dipolar field wouldn't be an issue and only the average value of the dipolar field across the pillar matters for SH switching. In the Supplementary Materials, we confirm that the composite external-field-free switching strategy still works for a $10 \times 10$ nm$^2$ circular pillar by performing micromagnetic simulations. This proves the scalability of the composite structure and promises its applications in high areal density memory arrays.

**Comparison with other field-free SH devices.**

As has been mentioned, previous field-free SH switching of PMA materials follows two paths: introducing a lateral structural asymmetry[20–25] or using a symmetry-breaking exchange field[26–32]. Table 1 compares our dipole-coupled SH device with the previous approaches. Among all the parameters, the in-plane compensation field $H_{comp}$ and out-of-plane shift field (per current density) $H_{shift} / J$ reflect the strength of symmetry-breaking effective field, for which our device has significantly greater values than most of the other two types of devices. This is reasonable because the local dipolar field is more likely to be strong than either the PMA gradient induced effective field or the in-plane exchange bias field in a PMA system. Additionally, the strength of the dipolar field is easier to be tuned via engineering the distance between PL and IPL, the saturation magnetization of IPL, or the shape of the nanopillar (see Supplementary Materials). This makes the dipole-coupled devices more practical than exchange-coupled devices, which requires high field post-annealing, suffers from the trade-off between PMA and $H_{ex}$, and is sensitive to Joule heating.

Although utilization of AFM as spin Hall material also enables the exchange-coupled SH devices to be integratible with p-MTJ[28–30,32], this strategy exclude the possibility of using various

SH materials that are not AFM but have a larger spin Hall efficiency such as topological insulators[4,11–14] Additionally, the analogue manner of exchange-coupled devices, attributed to the multiple domain states in the AFM layer, makes them more suitable to realize memristors for the application of neuromorphic computing[27,29,32] rather than memories. In contrast, the dipole-coupled devices, being compatible with a wide range of giant spin Hall materials and readily integratable with p-MTJ, would be more desirable for SOT based memory applications.

In summary, we have demonstrated an external-field-free SHE induced perpendicular magnetization switching by using a dipole-coupled composite structure, where an in-plane magnet can provide a sufficiently large biasing field to break the lateral symmetry. Robust full switching of nano-scale Ta/CoFeB/MgO pillar in a composite stack has been experimentally demonstrated with both DC and pulse currents in the absence of external field, with the critical current density of about $1.5 \times 10^7$ A/cm$^2$. Large in-plane compensation field of 135 Oe and out-of-plane loop shift of 45 Oe / $10^7$ A·cm$^{-2}$ have been obtained. The switching process of a $300 \times 150$ nm$^2$ nanopillar has been studied using micromagnetic simulations, taking the non-uniform stray field profile into considerations. The field-free switching of a $10 \times 10$ nm$^2$ composite pillar has also been demonstrated by simulation, confirming the scalability of the composite structure. Our dipole-coupled strategy, being compatible with various spin Hall systems and p-MTJ, pave the way towards energy-efficient SOT-based memory and logic applications with field-free operations.

## Methods

**Film preparation and characterization.** Samples were fabricated on thermally oxidized silicon substrates using DC sputtering in a Shamrock sputter tool. The deposited stack of the composite sample consists of Ta (5)/CoFeB (1.2)/MgO (2)/CoFeB (3)/MgO (2)/Ta (5) (from bottom to top, with thicknesses in nanometers), where the composition of CoFeB is $Co_{20}Fe_{60}B_{20}$; and the stack of the reference sample consists of Ta (5)/CoFeB (1.2)/MgO (2)/Ta (5). Layer thicknesses were chosen after careful optimization. The perpendicular anisotropy field is $H_k \approx 0.35$ T for the perpendicular CoFeB layer in both structures, determined from the difference between the in-plane and out-of-plane *M-H* loops.

**Device fabrication.** Using electron-beam lithography and ion milling procedures, both the composite sample and the reference sample were patterned into elliptical nanopillars on top of Ta Hall bars (Fig. 1b). The width of the current channel is 500 nm, and the width of the voltage branch is 200 nm. The elliptical pillars are with different aspect ratios including $285 \times 95$ nm$^2$ (3:1), $300 \times 120$ nm$^2$ (2.5:1), and $300 \times 150$ nm$^2$ (2:1). All experimental results presented in the main text are obtained from the $285 \times 95$ nm$^2$ device. The etching time of the pillar was carefully optimized to ensure that the CoFeB (1.2nm) layer was fully patterned. The sheet resistance of the remaining Ta layer after etching is around 640 $\Omega$ for both the composite and the reference samples.

**Electrical measurement setup.** A Keithley 6221 current source and a Keithley 2182A nanovolt metre were used in the Hall measurement. The external in-plane and out-of-plane magnetic fields were generated by coils driven by a Kepco power supply. In the pulse switching measurement (Fig. 4c), the current pulses were generated by Keithley 6221, and the Hall voltage were measured after the finishing of each pulse.

**Micromagnetic simulation.** The micromagnetic simulations were performed utilizing the Object Oriented Micro Magnetic Framework (OOMMF)[44]. The SOT induced by spin Hall effect is described by xf_stt extension, STT driven domain wall motion is described using Anv_SpinTEvolve extension[45], and Dzyaloshinskii–Moriya interaction is described by DMExchange6Ngbr extension. The parameters used in the simulations are: saturation magnetization $M_s = 1200$ emu/cm$^3$, exchange constant $A = 20 \times 10^{-12}$ J/m, uniaxial anisotropy field of PL $H_k = 1.8$ T, Gilbert damping $\alpha = 0.02$, spin polarization $P = 0.5$, non-adiabatic parameter $\beta = 0.04$, spin Hall angle $\theta_{SH} = 0.15$, Dzyaloshinskii–Moriya constant $|D| = 0.5$ mJ/m$^2$, and current density $J = 30 \times 10^{11}$ J/m$^2$. The geometry of the simulated block (in Fig. 5) is the same as what was patterned in the experiment, i.e. $300 \times 150$ nm$^2$ ellipses with the thickness of 1 nm for PL and 3 nm for IPL.

## Acknowledgements

This work was supported by C-SPIN, one of six centers of STARnet, a Semiconductor Research Corporation program, sponsored by MARCO and DARPA. Parts of this work were carried out at the University of Minnesota Nanofabrication Center, which receives partial support from NSF through the NNIN program, and the Characterization Facility, which is a member of the NSF-funded Materials Research Facilities Network via the MRSEC program.

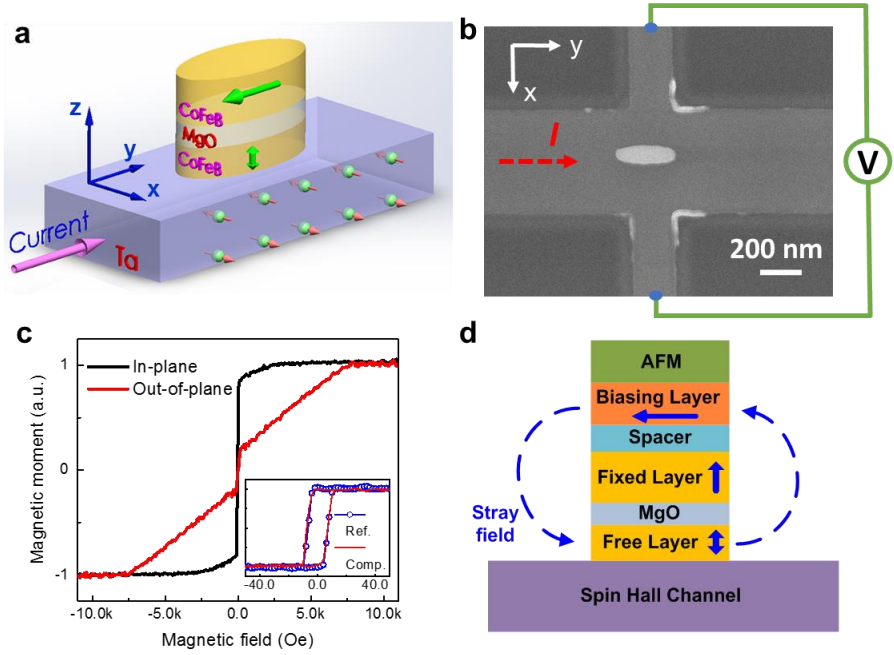

**Figure 1. Schematic of the composite nanopillar and magnetic properties.** (**a**) Schematic of the composite structure Ta(5)/CoFeB(1.2)/MgO(2)/CoFeB(3)/MgO, which is patterned into a elliptical nanopillar on top of Ta Hall bar. (**b**) Scanning electron micrograph of the fabricated device and setup of the measurement. The dimension of the elliptical pillar is $285 \times 95$ nm$^2$. (**c**) In-plane and out-of-plane *M-H* loop of the composite structure. Inset: out-of-plane *M-H* loop of the composite structure and the reference structure, respectively, in a narrow field range. (**d**) The proposed structure of a full p-MTJ stack for external-field-free SH switching utilizing the composite structure.

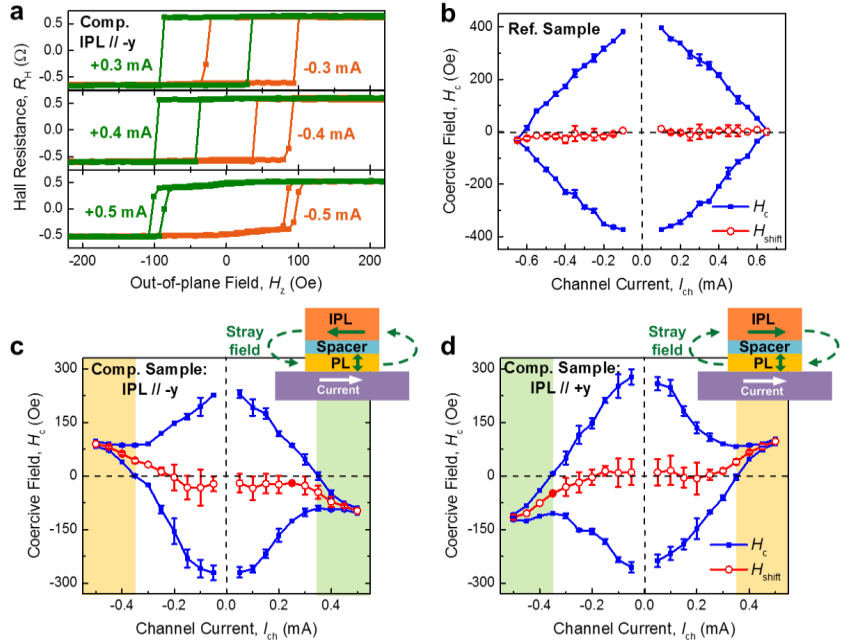

**Figure 2.** $R_H$-$H_z$ **loops measured with various** $I_{ch}$**.** (**a**) $R_H$-$H_z$ loops with different $I_{ch}$ for the composite sample, where IPL // −$y$. (**b**) Coercive field $H_c$ (square symbol) and shift of loop $H_{shift}$ (circular symbol) as a function of $I_{ch}$ for the reference sample. (**c**, **d**) $H_c$ and $H_{shift}$ as a function of $I_{ch}$ for the composite sample, with IPL been initialized to IPL // −$y$ in (**c**) and IPL // +$y$ in (**d**), respectively (the schematics of the magnetization configuration are shown in the top-right panel of (**c**) and (**d**)). The regions marked with green/yellow in (**c**) and (**d**) (i.e. $I_{ch}$ < −0.35 mA and $I_{ch}$ > +0.35 mA) are where $H_c^+$ and $H_c^-$ have the same sign, implying only one orientation of PL is stable in these regions.

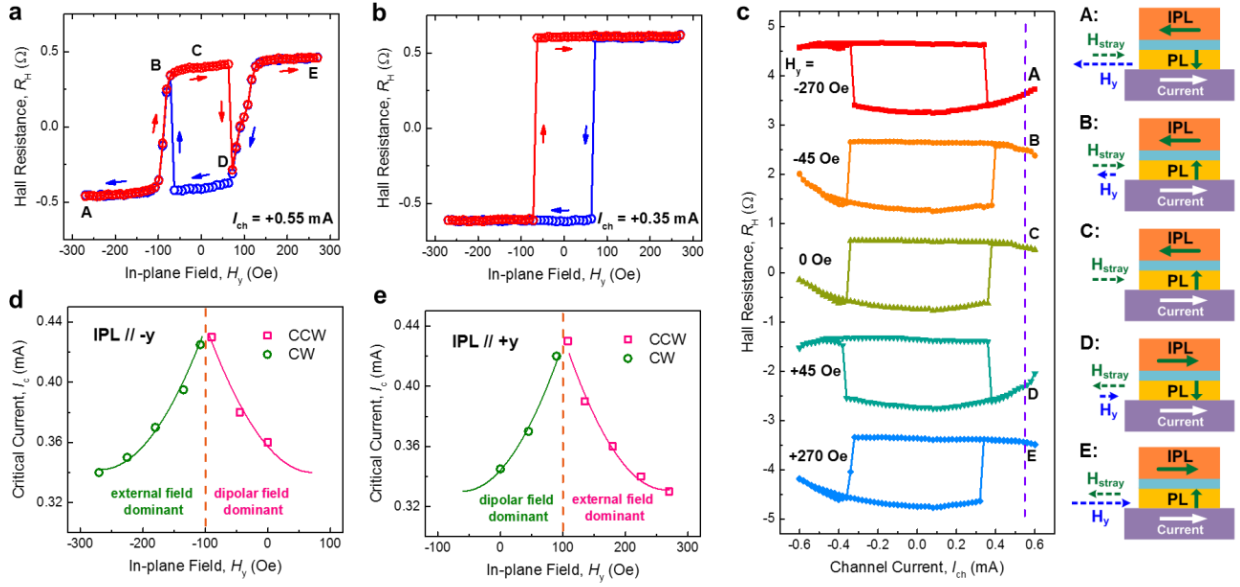

**Figure 3. Switching of PL nanomagnet measured with various external in-plane field $H_y$.** (**a, b**) $R_H$-$H_y$ loops measured with $I_{ch}$ = +0.55 mA in (**a**) and $I_{ch}$ = +0.35 mA in (**b**). (**c**) $R_H$-$I_{ch}$ loops under various $H_y$, which is gradually changed from −270 Oe to +270 Oe. Five points A-E representing five critical states of the device are labelled in both (**a**) and (**c**). Schematics on the right panel of (**c**) illustrate the magnetization configurations of the five points, where the lengths of the dash arrows represent the strengths of the fields. (**d, e**) Critical current $I_c$ as a function of $H_y$ extracted from the $R_H$-$I_{ch}$ loops, for IPL // −y in (**d**) and IPL // +y in (**e**), respectively. $I_c$ values obtained from clockwise (CW) and counter-clockwise (CCW) loops are plotted as different symbols. Solid lines are fitting to the data.

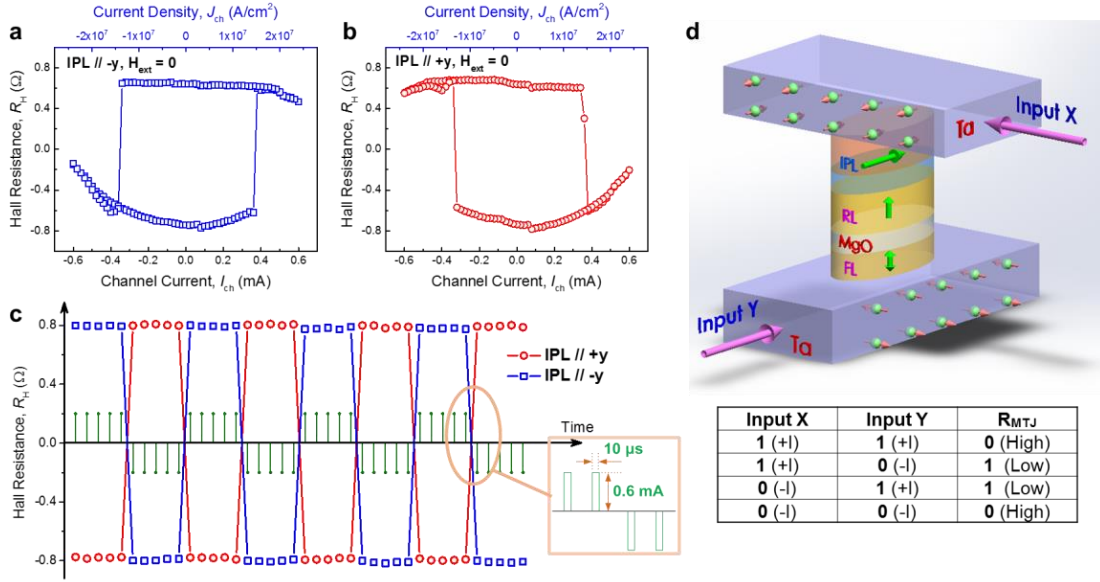

**Figure 4. Zero-field switching results.** (**a**, **b**) $R_H$-$I_{ch}$($J_{ch}$) loops measured without any external field. The IPL is initialized to IPL // −y in (**a**) and IPL // +y in (**b**), respectively. (**c**) $R_H$ variation upon application of a sequence of current pulses without external field. The pulse amplitude is 0.6 mA and pulse width is 10 µs. The configurations of IPL // −y and IPL // +y are plotted as different symbols. (**d**) Proposed XOR gate and its truth table based on the composite SH p-MTJ, where FL, RL and IPL denote free layer, reference layer and in-plane biasing layer, respectively.

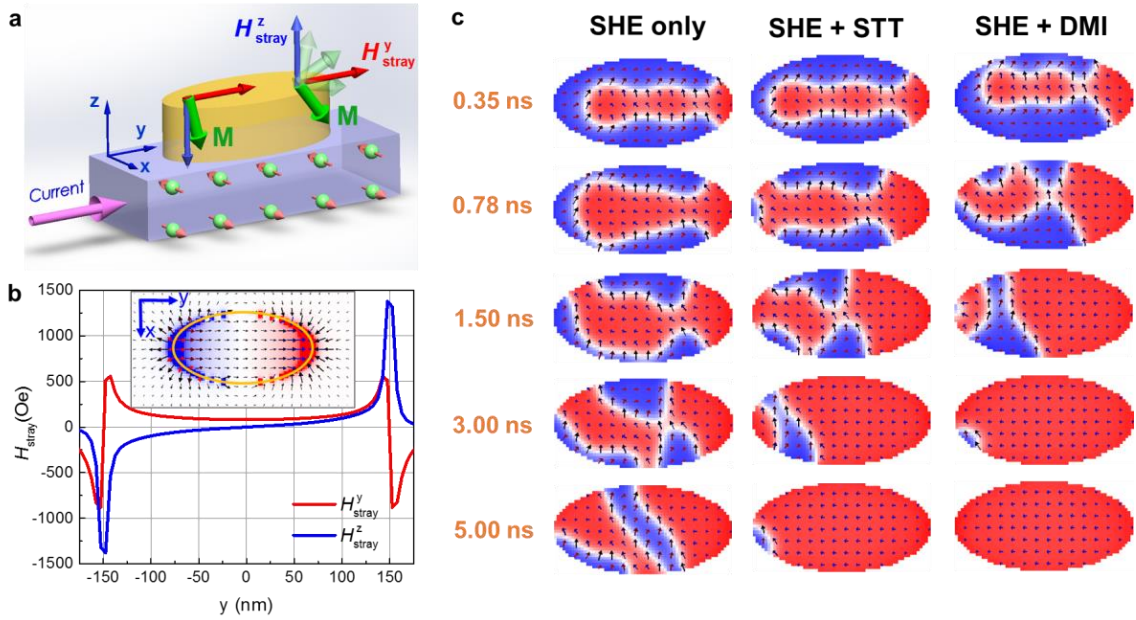

**Figure 5. Micromagnetic simulation results describing stray field distribution and possible switching process.** (**a**) Schematic showing the non-uniform stray fields from IPL // −y, and the resultant non-coherent switching of the magnetization. (**b**) Stray field profile along the long axis of PL for 300 × 150 nm$^2$ pillar. Inset: mapping of the stray field distribution on the PL, where the arrows show the strengths and directions of the in-plane component of the stray field, and color pixels show the strengths and directions of the z-component of the stray field (red = up, blue = down). (**c**) Simulated magnetization evolution in the PL, under the non-uniform stray field shown in (**b**) and a constant channel current along +y (the initial magnetization of PL is pointing down marked as blue color). Although the SHE alone cannot lead to a complete switching, the switching can be completed if involving STT or DMI driven domain wall motion. The parameters used in this simulation are in Methods.

**Table 1. Comparison of three types of external-field-free SH devices.**

| Mechanism | Ref. | Structure | Device Type | $H_k$ | $H_{comp}$ | $H_{shift}$ / J | $J_c$ |
|---|---|---|---|---|---|---|---|
| Tilted anisotropy / Anisotropy gradient | 20 | //Ta(5)/CoFeB(1)/TaO | Hall bar | 0.04 T ~0.12 T | n/a | 0~56 Oe / $10^7$ A·cm$^{-2}$ | 0.25 ×$10^7$ A·cm$^{-2}$ |
| | 23 | //Ta (10)/CoFeB (1)/MgO (1) | Nanopillar | 0.5 T | n/a | n/a | 2.5 ×$10^7$ A·cm$^{-2}$ |
| Exchange coupling | 26 | //Ta(1)/Pt(5)/CoFe(0.8)/Ru(2)/CoFe (1.5)/IrMn(10)/Pt(2) | Hall bar | n/a | 400 Oe (0.8T annealing) | 22 Oe / $10^7$ A·cm$^{-2}$ | 3×$10^7$ A·cm$^{-2}$ |
| | 27 | //Ta(1)/Pt (3)/Co (0.7)/Pt (0.3)/IrMn (6)/TaOx (1.5) | Hall bar | 1T | 50 Oe (2T annealing) | n/a | 8×$10^7$ A·cm$^{-2}$ |
| | 28 | //Ta(5)/CoFeB(3)/IrMn(3)/CoFeB(1)/MgO(1.6)/Ta(2) | Hall bar | 0.5 T | 50 Oe (0.8T annealing) | n/a | 4.2×$10^7$ A·cm$^{-2}$ |
| | 29 | //PtMn(8)/[Co(0.3)/Ni(0.6)]$_2$/MgO | Hall bar | 0.15 T | 83 Oe (1.2T annealing) | 28 Oe / $10^7$ A·cm$^{-2}$ | 0.6×$10^7$ A·cm$^{-2}$ |
| | 31 | //Ta(2)/Pt(3)/CoFe(1.1)/IrMn (3)/Pt(1) | Hall bar | 1 T | 95 ~ 215 Oe (1.5T annealing) | n/a | 3.1×$10^7$ A·cm$^{-2}$ |
| | 32 | //Ta(5)/PtMn(10)/CoFeB/Gd/CoFeB/MgO | Hall bar | 0.3 T | 22 Oe (0.5T annealing) | 10 Oe / $10^7$ A·cm$^{-2}$ | 0.96×$10^7$ A·cm$^{-2}$ |
| Dipole coupling | **This work** | //Ta(5)/CoFeB(1.2)/MgO(2)/CoFeB (3)/MgO(2)/Ta | Nanopillar | 0.35 T | 100 ~ 135 Oe | 45 Oe / $10^7$ A·cm$^{-2}$ | 1.5×$10^7$ A·cm$^{-2}$ |

Supplementary Materials for

# External-Field-Free Spin Hall Switching of Perpendicular Magnetic Nanopillar with a Dipole-Coupled Composite Structure

Zhengyang Zhao[†], Angeline K. Smith[†], Mahdi Jamali, and Jian-Ping Wang[*]

Department of Electrical and Computer Engineering, University of Minnesota, 4-174 200 Union Street SE, Minneapolis, MN 55455, USA

**This file includes:**

- Section S1. Results for composite nanopillars with IPL // x.
- Section S2. Results for composite nanopillars with different aspect ratios.
- Section S3. Coercive field and thermal stability of IPL.
- Section S4. Calculation of stray field for $300 \times 150$ nm$^2$ nanopillar.
- Section S5. Micromagnetic simulation of SH switching process for $300 \times 150$ nm$^2$ elliptical nanopillar under a uniform in-plane field.
- Section S6. Micromagnetic simulation of external-field-free SH switching for $10 \times 10$ nm$^2$ composite nanopillar

## Section S1. Results for composite nanopillars with IPL // x

As described in the main text, a successful switching requires the easy axis of IPL to be along y-axis (as shown in Fig. 1 in the main text), so that the dipolar field from IPL is mainly along y-direction and can substitute the externally applied field $H_y$. This concept has been proved by comparing the composite device with an unbiased reference device (Fig. 2b in the main text). Here we show another device for comparison, which was patterned from the same composite stack but with nanopillar's long axis along x-axis (in the transverse direction). Then the IPL magnetization would point to +x or –x direction due to the shape anisotropy. Figure S1 shows the result for one of such devices, by performing the same measurement as shown in Fig. 2 in the main text. As can be seen, the curves under positive $I_{ch}$ and negative $I_{ch}$ are symmetric and $H_{shift}$ is zero at a large current, unlike the symmetry-breaking behavior in Fig. 2 c-d in the main text. The non-zero shift under a small current in Fig. S1, as also appeared in Fig. 2 c-d in the main text, is probably due to the existence of a perpendicular component in the magnetization of IPL. Additionally, in the current sweeping measurement (not shown), the transverse composite pillar didn't shown deterministic switching. This is in consistence with the fact that an applied field $H_x$ along the transverse direction would not result in the SH switching of a PMA layer.

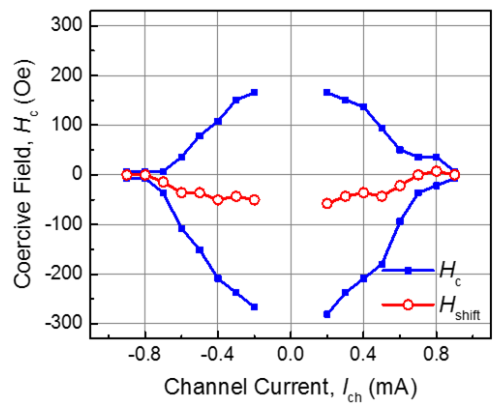

**Figure S1. Variation of $H_c$ and $H_{shift}$ versus $I_{ch}$ for composite nanopillars with its long axis along x-direction.** Coercive field $H_c$ and loop shift $H_{shift}$ as a function of $I_{ch}$, extracted from the $R_H$-$H_z$ loops, for the composite device with an elliptical pillar of $95 \times 285$ nm$^2$, where its long axis is along x-direction (transverse direction). The width of channel is 1 μm.

## Section S2. Results for composite nanopillars with different aspect ratios

The results obtained from the composite nanopillar with the dimension of $285 \times 95$ nm$^2$ (3:1) are shown in the main text. Here we present the results of devices with other aspect ratios, i.e. $300 \times 120$ nm$^2$ (2.5:1) pillar and $300 \times 150$ nm$^2$ (2:1) pillar, in Fig. S2 and Fig. S3. In Fig. S2, the curve shows $H_c$ and $H_{shift}$ versus $I_{ch}$ for $300 \times 120$ nm$^2$ and $300 \times 150$ nm$^2$ composite pillars. The two-fold symmetry behavior of the curve is similar to that shown in Fig. 2 in the main text, and the amplitude of $H_{shift}$ is close to that of the 3:1 pillar (shift around 100 Oe at $I_{ch}$ = 0.5 mA). This indicates the pillar is not necessarily to have a large aspect ratio in order to be switchable without an external field. With the IPL been pinned, the aspect ratio of the pillar could be decreased to 1. The reason of making elliptical pillars with a large aspect ratio in this paper is to guarantee the magnetization of IPL is thermally stable, since we don't get the IPL pinned. As will be seen in the next section, the thermal stability of IPL decreases dramatically with the aspect ratio approaching to 1.

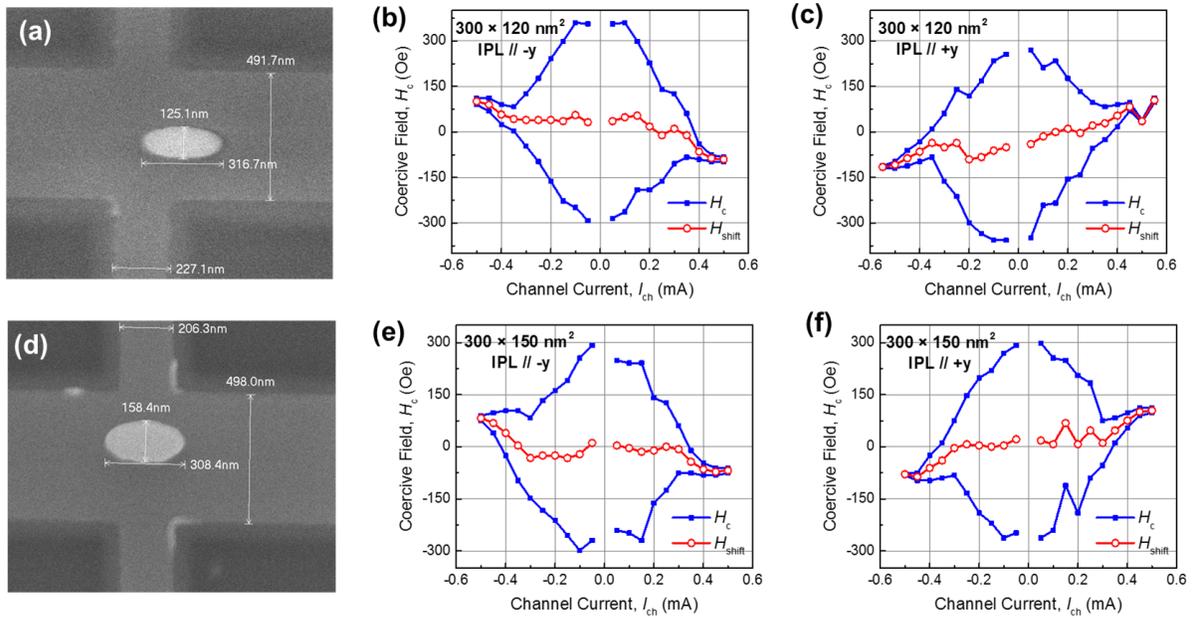

**Figure S2. Variation of $H_c$ and $H_{shift}$ versus $I_{ch}$ for composite pillars with different aspect ratio.** (**a**), (**d**) SEM images of the device (**a**) with $300 \times 120$ nm$^2$ (2.5:1) pillar and (**d**) with $300 \times 150$ nm$^2$ (2:1) pillar. (**b**),

(**c**) Coercive field $H_c$ and loop shift $H_{shift}$ as a function of $I_{ch}$ for the $300 \times 120$ nm² composite device, extracted from the $R_H$-$H_z$ loops, with IPL been initialized to IPL // −y in (**b**) and IPL // +y in (**c**), respectively. (**e**), (**f**) $H_c$ and $H_{shift}$ as a function of $I_{ch}$ for the $300 \times 150$ nm² composite device, with IPL been initialized to IPL // −y in (**e**) and IPL // +y in (**f**), respectively. The result for $285 \times 95$ nm² (3:1) composite device is shown in **Fig. 2** (**c-d**) in the main text.

Figure S3 plots the critical current $I_c$ as a function of $H_y$ and the boundary between CCW and CW switching loops for $300 \times 120$ nm² and $300 \times 150$ nm² composite pillars, by performing the same measurement as in Fig. 3 in the main text. As can be seen, with the aspect ratio approaching 1, the compensation field $H_{comp}$ increases from 100 Oe (3:1 pillar) to 108 Oe (2.5:1 pillar) to 135 Oe (2:1 pillar). This is due to the increasing of the stray field in the y-direction, as can be seen from the calculated result in Fig. S3c.

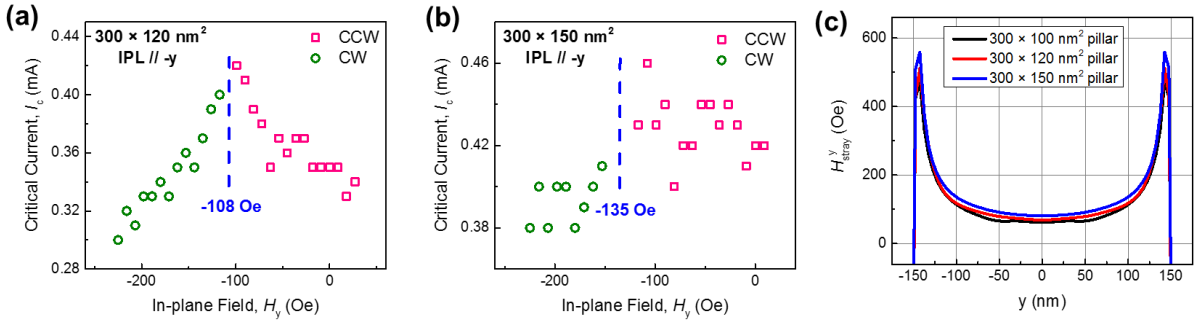

**Figure S3. Critical current $I_c$ versus external field $H_y$ for composite pillars with different aspect ratio.** (**a**) $I_c$ versus $H_y$ for the composite device with $300 \times 120$ nm² (2.5:1) pillar. (**b**) $I_c$ versus $H_y$ for the composite device with $300 \times 150$ nm² (2:1) pillar. The directions of IPL are all along –y. The result for $285 \times 95$ nm² (3:1) composite device is shown in **Fig. 3** (**d-e**) in the main text. (**c**) The calculated stray field $H_{stray}^y$ along the long axis of PL, generated from the IPL with different aspect ratios.

## Section S3. Coercive field and thermal stability of IPL

In the main text, we have mentioned the switching of PL between C and D in Fig. 3a is because IPL is reversed by the external field $H_y$. Apparently, the coercive field of IPL $H_c^{IPL}$ depends on both the channel current (Joule heating) and the aspect ratio of the elliptical pillar (shape anisotropy), as shown in Fig. S4a. $H_c^{IPL}$ was determined by consecutively detecting the magnetization direction of IPL after applying $H_y$ with a series of different strengths; and the direction of IPL was determined from the polarity of the switching loop of PL.

The thermal stability of IPL is essential for a successful external-field-free switching. As shown in Fig. S4b-c, with the pulse current increases from 0.45 mA to 0.53 mA, the error rate for the switching of a $300 \times 150$ nm$^2$ nanopillar increases significantly. This is because the stray field generated from IPL becomes weak due to the thermal fluctuation of the in-plane magnetization under a large current. Additionally, with the same current pulse, the error rate for $285 \times 95$ nm$^2$ nanopillar is much lower (see Fig. 4c in the main text), because of its stronger shape anisotropy. For practical memory application, the thermal stability of IPL can be enhanced with a pinning layer instead of a large aspect ratio.

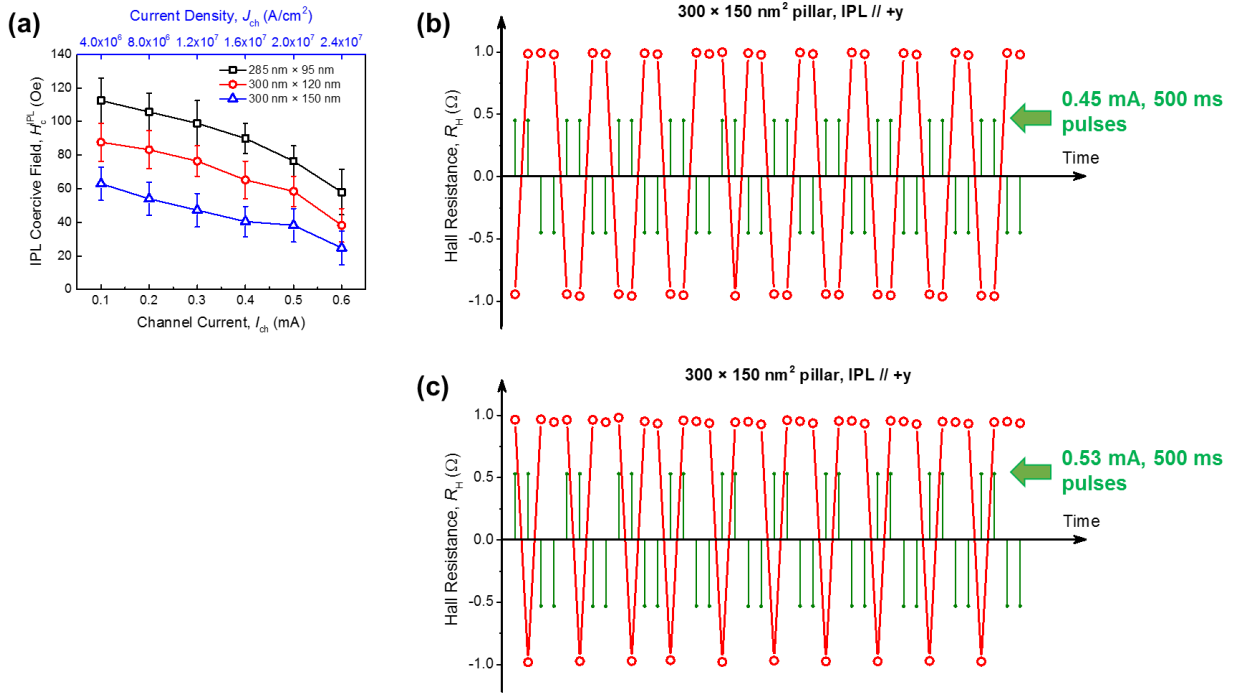

**Figure S4. Coercive field of IPL versus channel current and the effect of IPL's thermal stability on the switching probability.** (**a**) The coercive field of IPL $H_c^{IPL}$ as a function of various channel current $I_{ch}$ applied. The three curves are for devices with $285 \times 95$ nm$^2$ (3:1), $300 \times 120$ nm$^2$ (2.5:1) and $300 \times 150$ nm$^2$ (2:1) pillars, respectively. The channel width of all devices are 500 nm. (**b**), (**c**) $R_H$ variation upon application of a sequence of current pulses without external field for the composite device with $300 \times 150$ nm$^2$ (2:1) pillars. In (**b**) the pulse amplitude is 0.45 mA and pulse width is 500 ms, in (**c**) the pulse amplitude is 0.53 mA and pulse width is 500 ms. The directions of IPL are all along +y.

## Section S4. Calculation of stray field for 300 × 150 nm² nanopillar

The strength and profile of the stray field generated from IPL depends on the saturation magnetization of IPL, the thickness of IPL, the distance from PL to IPL, and the dimension of the nanopillar. This provide flexibility in engineering the composite structures for various applications. Figure S5 shows the stray field profiles with different IPL thicknesses and different distances for 300 × 150 nm² elliptical pillar.

It can be seen that the stray field decays fast out of the region of the elliptical pillar. Therefore, the stray field from one cell will have little influence on its neighbor cells in a memory array.

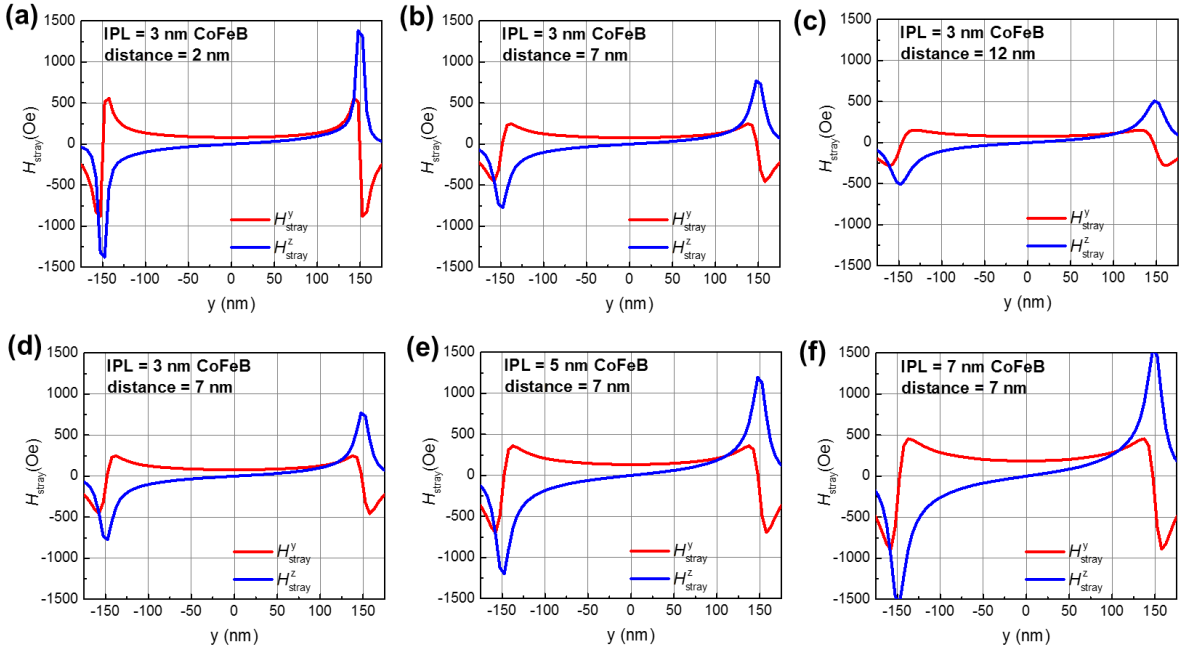

**Figure S5. Calculated stray field depending on IPL thickness and distance between PL and IPL for 300 × 150 nm² elliptical pillar.** (**a-c**) IPL's stray field profile along the long axis of PL layer, with the same IPL thickness of 3 nm and the different distances between PL and IPL. (**d-f**) IPL's stray field profile along the long axis of PL layer, with different IPL thickness and same distances between PL and IPL.

## Section S5. Micromagnetic simulation of SH switching process for 300 × 150 nm² elliptical nanopillar under a uniform in-plane field

Figure S6 present the micromagnetic simulation of SH switching of a normal Ta/CoFeB/MgO PMA structure under a uniform external in-plane field of $H_y$ = 500 Oe. It can be seen the reversal starts from the center of the elliptical pillar, because the local demagnetization field is stronger at the center than at the edge of the pillar.

Compared with Fig. S6, the switching process under a non-uniform dipolar field shown in Fig. 5c in the main text is quite different. For the composite structure, the reversed domain first nucleates from one end of the pillar, and then propagates to the rest area (Fig. 5c in the main text).

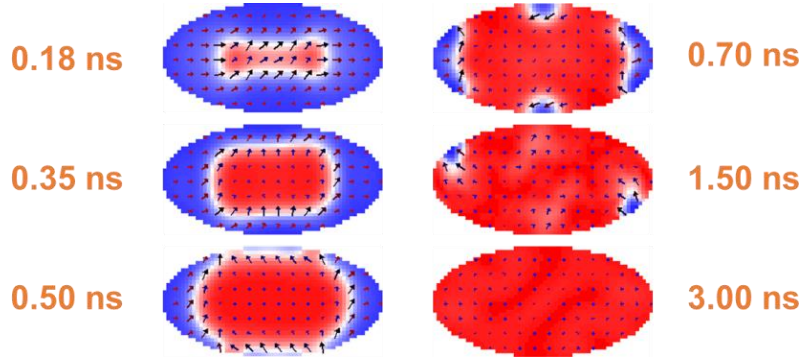

**Figure S6. Micromagnetic simulation showing the SH switching process of a normal PMA structure under an external in-plane field.** The dimension of the elliptical pillar is $300 \times 150$ nm$^2$. Only SHE is considered (without involving STT or DMI). The applied uniform field is $H_y = 500$ Oe, the current density is $J = 20 \times 10^{11}$ J/m$^2$, and other parameters are the same as the simulation results presented in Fig. 5 in the main text.

## Section S6. Micromagnetic simulation of external-field-free SH switching for $10 \times 10$ nm$^2$ composite nanopillar

In this section, we show that the composite field-free switching strategy still works for a $10 \times 10$ nm$^2$ circular pillar by performing a micromagnetic simulation. Figure S7 (**a**-**b**) shows the profile of stray field across PL in a $10 \times 10$ nm$^2$ composite nanopillar, where the distance between IPL and PL is set to 7 nm to mimic the distance in full MTJ stacks. Large IPL saturation magnetization of 1700 emu/cm$^3$ and IPL thickness of 7 nm are used to enhance the stray field. The magnetization direction of IPL is set to $-y$ direction assuming it has been pinned by a top anti-ferromagnetic layer. Other simulation parameters are shown in Table S1. It can be seen that $H_{\text{stray}}^y$ gets its maximum at the center of the pillar, different from the case of a 300 × 150 nm² pillar where $H_{stray}^y$ gets its maximum at the both ends of the long axis.

For nanopillars smaller than 40 nm, the magnetization switching mode would be single domain type rather than nucleation type[36]. Therefore, the non-uniform profile of the stray field would no longer be an issue for 10 × 10 nm² nanomagnet as it is for the 300 × 150 nm² pillar. In Fig. S7 (**c**), the solid curve shows the successful SH switching of the 10 × 10 nm² nanomagnet only with the assistance of the stray field. For comparison, the dash curve shows the SH switching of the same nanomagnet but only with a uniform external field of $H_y$ = 550 Oe. The two trajectories are quite close to each other, indicating the effect of the stray field is similar with the effect of a uniform external field on the PL nanomagnet. Since the nanomagnet switches as a single domain, only the average value of the dipolar field across the pillar matters for SH switching. According to Fig. S7 (**b**), the average value of $H_{stray}^y$ across the pillar is around 500 Oe, in agreement with Fig. S7 (**c**). These results prove the scalability of the composite structure and promise its applications in high areal density memory arrays.

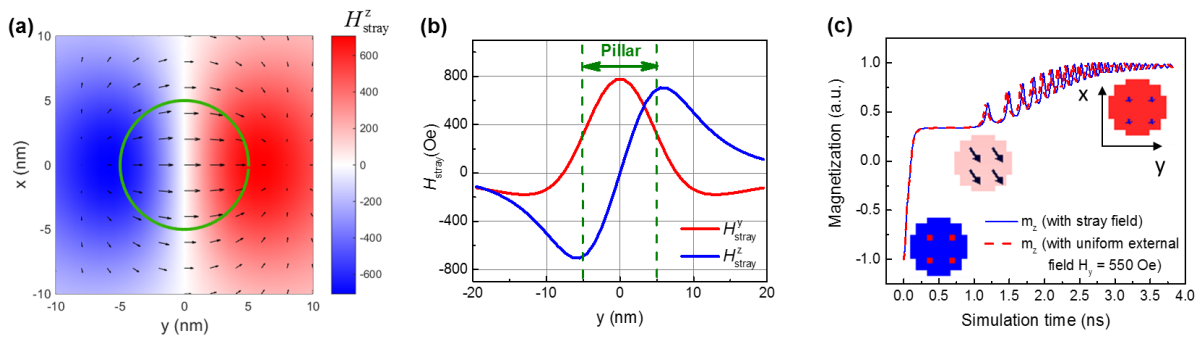

**Figure S7. Micromagnetic simulation showing the external-field-free SH switching of a 10 × 10 nm² nanopillar.** (**a**) Mapping of the stray field distribution across PL for 10 × 10 nm² circular pillar, where the arrows show the strengths and directions of the in-plane component of the stray field, and color pixels show the strengths and directions of the z-component of the stray field (red = up, blue = down). (**b**) Stray field profile along *y*-axis across PL. (**c**) Switching curves of PL with current density of $J = 20 \times 10^{11}$ J/m², where the solid line represents SH switching under the stray field of IPL and the dash line represents the switching under an uniform external field of $H_y$ = 550 Oe. Inset: top-view of PL magnetization pattern of initial state, intermediate state and final state. Parameters of the simulation is shown in Table S1.

**Table S1. Simulation parameters for SH switching of 10 × 10 nm² composite nanopillar**

| | |
|---|---|
| IPL dimensions | 10 nm × 10 nm × 7 nm |
| PL dimensions | 10 nm × 10 nm × 1 nm |
| Cell size | 1 nm × 1 nm × 0.5 nm |
| Distance between PL and IPL | 7 nm |
| | |
| IPL saturation magnetization | 1700 emu/cm³ |
| PL saturation magnetization | 1200 emu/cm³ |
| Exchange constant | $20 \times 10^{-12}$ J/m |
| Gilbert damping | 0.02 |
| PL perpendicular anisotropy | 1.5 T |
| Spin Hall angle | 0.15 |
| Current density | $20 \times 10^{11}$ J/m² |